\documentclass[12pt,epsfig]{article}
\usepackage{etex}
\usepackage{epsfig}
\usepackage{epsf,latexsym}
\usepackage{amsfonts,amssymb,amsmath} 
\usepackage{mathtools}
\usepackage{tikz}
\epsfverbosetrue
\newcommand{\double}[1]{\mathbb{#1}}

\newcommand{\rr}{\double{R}}

\newcommand{\llll}{\mathcal{L}}

\newcommand{\dee}{\hbox{\rm{D}}}
\newcommand{\de}{\hbox{\rm{d}}}

\newcommand{\pa}{\partial}

\newcommand{\dpp}{\vcentcolon}
\newcommand{\bb}{\begin{eqnarray}}
\newcommand{\ee}{\end{eqnarray}}
\newcommand{\eee}{\nonumber\end{eqnarray}}
\newcommand{\qq}{\quad}

\begin{document}

\font\twelve=cmbx10 at 13pt
\font\eightrm=cmr8

\thispagestyle{empty}

\begin{center}
${}$
\vspace{3cm}

{\Large\textbf{Torsion, an alternative to the cosmological constant?}} \\

\vspace{2cm}

{\large

Thomas Sch\"ucker\footnote{thomas.schucker@gmail.com } (CPT\footnote{Centre de Physique Th\'eorique, 
Aix-Marseille Univ; CNRS, UMR 7332; Univ Toulon;\\
13288 Marseille Cedex 9, France}),
Andr\'e Tilquin\footnote{tilquin@cppm.in2p3.fr }
(CPPM\footnote{Centre de Physique des Particules de
Marseille,
Aix-Marseille Univ; CNRS, UMR 6550;\\
13288 Marseille Cedex 9, France})}

\vspace{3cm}

{\large\textbf{Abstract}}
\end{center}
We confront Einstein-Cartan's theory with the Hubble diagram and obtain a negative answer to the question in the title. Contrary findings in the literature seem to stem from an error in the field equations.
\vspace{3.2cm}

\vskip 2 truecm

\noindent CPT-2011/P010\\
\noindent 1109.4568

\section{Introduction}

A recent fit \cite{ts} of Einstein-Cartan's theory \cite{car,hehl76,tworecent}  to the Hubble diagram of supernovae was rather encouraging in that  parity conserving spin-density as the source of torsion, could -- within today's error-bars -- replace cold dark matter. The underlying space was flat, homogeneous, isotropic
and invariant under inversion. In the present work we extend our fit by also admitting the possible parity violating term. Our motivation is that weak forces do break parity. The additional term comes with a new parameter (called $w_{\tilde s}$ in this paper). In \cite{ts} our goal was not to increase the number of parameters: the parity even parameter $w_s$ was introduced in order to set to zero the cold dark matter parameter. We maintain this goal here and we try to set the cosmological constant to zero when admitting $w_{\tilde s}$. We find that in spite the large present day error bars, this hypothesis is excluded by the Hubble diagram of supernovae.

Parity odd Einstein-Cartan theory in the context of maximally symmetric cosmology has a long history. Already in 1978, Bloomer \cite{bloo} analyzed the theory, with the parity odd part only, on the 3-sphere. In 1986, Peter Minkowski \cite{mink} reconsidered it on flat $\rr^3$. He mentions the handedness of spiral galaxies as motivation. More recently, in 2002
Capozziello {\it et al.}  \cite{torLam} took up the flat, parity odd theory again and found that spin-density parameterized by $w_{\tilde s}$ (in our notations) can replace the cosmological constant. However they miss a factor 3 in the field equations. Our analysis shows that their result does not hold after correction.

\section{Notations and field equations}

We use the conventions of reference \cite{ts}. For the reader's convenience we briefly summarize them.

Let $x^\mu $ ($\mu = 0,1,2,3$) be a coordinate system on an open subset of $\rr^4$. We will use both a holonomic frame $\de x^\mu$ and an orthonormal frame, {\it un rep\`ere mobile} using Cartan's words, $e^a=\dpp {e^a}_\mu \,\de x^\mu,\ a= 0,1,2,3 $. We denote a metric connection with respect to an orthonormal frame by  ${\omega ^a}_b=\dpp{\omega ^a}_{b c }\,e^c $. It is as a 1-form with values in the Lie algebra of the Lorentz group. We follow the traditional convention and denote the same connection with respect to the holonomic frame by a different letter: ${\Gamma^\alpha  }_\beta  =\dpp {\Gamma^\alpha  }_{\beta \mu }\,\de x^\mu $, a $g\ell(4)$ valued 1-form. The link between the components of the connection with respect to the holonomic frame $\Gamma $ and with respect to the orthonormal frame $\omega $ is given by the $GL(4)$ gauge transformation with $e(x)={e^a}_\mu(x)\,\in GL(4)$;
\bb \omega =e\Gamma e^{-1}+e\de e^{-1}.\ee

 Then (suppressing all wedge symbols) Cartan's two structure equations read:
\bb R\dpp=\de \omega +{\textstyle\frac{1}{2}} [\omega ,\omega ], \label{cartan1}\ee
for the curvature 2-form  ${R ^a}_b=\dpp{\textstyle\frac{1}{2}} {R ^a}_{b cd }e^ce^d$, and
\bb T\dpp=\dee e=\de e+\omega e,\label{cartan2}\ee
for the torsion 2-form, ${T ^a}=\dpp{\textstyle\frac{1}{2}} {T ^a}_{ bc }e^be^c$.
It will be useful to decompose the torsion tensor  into its three irreducible parts: 
\bb T_{abc}=A_{abc}+\eta_{ab}V_c-\eta_{ac}V_b+M_{abc},\ee
with the completely antisymmetric part $A_{abc}\dpp={\textstyle\frac{1}{3}}(T_{abc}+T_{cab}+T_{bca})$, the vector part $V_c\dpp={\textstyle\frac{1}{3}} T_{abc}\eta^{ab}$, and the mixed part $M_{abc}$ characterized by $M_{abc}=-M_{acb}$, $M_{abc}\eta^{ab}=0$, and $M_{abc}+M_{cab}+M_{bca}=0$.

In a Riemann-Cartan space there are two kinds of geodesics: curves that minimize the arc-length (or proper time) with respect to the metric and curves whose tangent vectors are parallel with respect to the connection. These two geodesics coincide if and only if the torsion only has a completely antisymmetric part, $V=0,\ M=0$.

With these notations, the Einstein-Hilbert action reads
\bb S_{\rm EH}[e,\omega ]&=&\,\frac{-1}{32\pi G}\, 
\int (R^{ab}+{\textstyle\frac{1}{6}}\Lambda e^ae^b )\, e^ce^d\epsilon_{abcd}\nonumber\\
&=&
\,\frac{-1}{16\pi G}\, 
\int ({R^{ab}}_{ab}+2\Lambda )\,\de V,
  \ee
where  $\epsilon_{0123}=1$. 
The energy-momentum current is the vector-valued 3-form $\tau_a$ obtained by varying the orthonormal frame in the matter Lagrangian:
\bb \llll_{\rm M}[e+f,\omega ]-\llll_{\rm M}[e,\omega ]=\dpp -f^a\tau_a+O(f^2).\ee
The energy-momentum tensor $\tau_{ab}$ is defined by $\ast \tau_a=\dpp \tau_{ab}e^b$, where $\ast$ is the Hodge star of the metric.

Let us prove that the energy-momentum tensor is symmetric if torsion vanishes:
The curvature components ${R^{ab}}_{cd}$ are antisymmetric in $ab$ and $cd$. If torsion is zero, there is a third, cyclic symmetry:
\bb {R^a}_{bcd}+{R^a}_{cdb}+{R^a}_{dbc}=0. \ee
It is easily derived by applying the hodge star to the second Bianchi identity $\dee T= \dee\dee e=Re$.
Together with the other two symmetries, the cyclic symmetry implies that the Einstein tensor $G_{ck}\dpp={R^a}_{cak}-{\textstyle\frac{1}{2}} {R^{ab}}_{ab}\,\eta_{ck}$ is symmetric: $G_{ck}=G_{kc}$. Einstein's equation then tells us that we can consistently couple only matter with symmetric energy-momentum tensors. Since we have assumed vanishing torsion, this matter cannot depend on the connection $\omega$. 

We now show that Lorentz invariance of the matter action is sufficient to guarantee the symmetry of the energy-momentum tensor as long as the matter fields satisfy their proper field equations, i.e. as long as they are `on shell'. We recall that Lorentz invariance follows from the description of the metric by means of an orthonormal frame.

To calculate the variation of the matter Lagrangian under an infinitesimal Lorentz transformation $({\Omega^a}_b)\in so(1,3)$ we can immediately use the defining equation of energy-momentum (6) with 
\bb f^a=-{\Omega^a}_b\,e^b,\ee
because on shell the variation of the matter fields does not contribute:
\bb 0= {\Omega^c}_b\,e^b\, {\textstyle\frac{1}{6}} {\tau_c}^a \epsilon_{arsd} e^r e^s e^d. \ee
Applying the Hodge star we find:
\bb 0={\Omega^c}_b\, {\textstyle\frac{1}{6}} {\tau_c}^a \epsilon_{arsd} \epsilon^{brsd} =
-\Omega^{cb}\,\tau_{cb}. \ee
$\Omega^{cb}$ being an arbitrary antisymmetric matrix, this equation implies the symmetry of $\tau_{cb}$.

Likewise, the spin current is the Lorentz-valued 3-form $S_{ab}$ obtained by varying the connection in the matter Lagrangian:
\bb \llll_{\rm M}[e,\omega+\chi  ]-\llll_{\rm M}[e,\omega ]=\dpp -{\textstyle\frac{1}{2}} \chi ^{ab}S_{ab}+O(\chi ^2).\ee
The spin tensor $S_{abc}$ is defined by $\ast S_{ab}=\dpp S_{abc}e^c$.
Einstein's equations are obtained by varying the total action with respect to the orthonormal frame:
\bb
(R^{ab}+{\textstyle\frac{1}{3}}\Lambda e^ae^b)\,e^d\epsilon_{abcd} =-16\pi G\tau_c \qq {\rm or\  equivalently}\qq
 G_{ab} -\Lambda \eta_{ab} =8\pi G\tau_{ba}.\ee

Likewise Cartan's equations are derived by varying the total action with respect to the connection:
\bb T^ce^d\epsilon_{abcd}=-8\pi GS_{ab},
\ee
or equivalently:
\bb A_{cab}+2V_a\eta_{bc}-2V_b\eta_{ac}+M_{cab}=-8\pi GS_{abc}.\ee

\section{Homogeneous and isotropic spaces}

The invariance of the
 metric tensor $g_{\mu \nu}(x) = {e^a}_\mu(x)\,{e^b}_\nu(x)\,\eta_{ab}$ under an infinitesimal diffeomorphism $\xi $ is expressed by
 the Killing equation:
\bb 
\xi ^\alpha \,\frac{\pa}{\pa x^\alpha }\, g_{\mu \nu}+\,\frac{\pa \xi ^{\bar\mu }}{\pa x^\mu }\, g_{\bar\mu \nu}+\,\frac{\pa \xi ^{\bar\nu }}{\pa x^\nu }\, g_{\mu \bar\nu}=0.\label{kill}\ee
 Likewise the vector field $\xi $ preserves the connection if
\bb 
\xi ^\alpha \,\frac{\pa}{\pa x^\alpha }\,  {\Gamma ^\lambda }_{\mu \nu}
-\,\frac{\pa \xi ^{\lambda  }}{\pa x^{\bar\lambda } }\,  {\Gamma ^{\bar\lambda }}_{\mu \nu}
+\,\frac{\pa \xi ^{\bar\mu }}{\pa x^\mu }\,  {\Gamma ^\lambda }_{\bar\mu \nu}
+\,\frac{\pa \xi ^{\bar\nu }}{\pa x^\nu }\,  {\Gamma ^\lambda }_{\mu \bar\nu}
+\,\frac{\pa ^2\xi ^\lambda }{\pa x^\mu \pa x^\nu}\, 
=0.\label{kill2}\ee

The most general Riemann-Cartan space invariant under $SO(3)\ltimes\rr^3$ has the Robertson-Walker metric: $e^0=\de t,\, e^1=a\,\de x,\,e^2=a\,\de y,\,e^3=a\,\de z$, with the scale factor $a(t)$, a positive function of cosmic time $t$. The non-vanishing components ${\omega ^a}_{bc }$ of the most general $SO(3)\ltimes \rr^3$ invariant connection are:
\bb {\omega ^0}_{ij}={\omega ^i}_{0j}=\,\frac{b}{a}\, \delta _{ij},\qq {\omega ^i}_{jk}=\,\frac{f}{a}\, \epsilon_{ijk},\ee
with two additional functions $b(t)$ and $f(t)$. The first is parity even like the scale factor, the second is parity odd.

The Riemann tensor has the following non-vanishing components:
\bb 
{R^0}_{i0k}={R^i}_{00k}=\,\frac{b'}{a}\,\delta _{ik},&&
{R^0}_{ijk}=-2\,\frac{bf}{a^2}\,\epsilon_{ijk},\\
{R^i}_{j0k}=\frac{f'}{a}\, \epsilon_{ijk},&&
{R^i}_{jk\ell}=\frac{b^2-f^2}{a^2}\,(\delta _{ik}\delta _{j\ell}- \delta _{i\ell}\delta _{jk}).\ee
The Einstein tensor has:
\bb G_{00}=3\,\frac{b^2-f^2}{a^2}\,,\qq
G_{ij}=-\left(2\,\frac{b'}{a}\, +\,\frac{b^2-f^2}{a^2}\,\right)\,\delta _{ij}.\ee 
The torsion tensor has:
\bb
{T^i}_{0j}=\,\frac{a'-b}{a}\,\delta _{ij},\qq
 {T}_{ijk}=2\,\frac{f}{a}\,\epsilon_{ijk}.\ee
The antisymmetric part has only space components, $A_{ijk}=2f/a\,\epsilon_{ijk}$, the vector part has only a time component, $V_0=(b-a')/a$, and the mixed part vanishes, $M=0$.
This result agrees with the curvature and torsion found in references \cite{symt,gmh} for spacetimes with maximally symmetric 3-spaces.

\section{Equations of state and Friedmann equations}

The most general $SO(3)\ltimes\rr^3$-invariant
energy-momentum tensor contains two function of time, the energy density $\rho (t)$ and  the pressure $p(t)$ and one usually assumes an equation of state $p(t)=\dpp w\,\rho (t)$. 
 
 Likewise the most general $SO(3)\ltimes\rr^3$-invariant spin density has two functions of time $s(t)$ and $\tilde s(t)$ 
in the two irreducible components:
  $S_{0jk}=\dpp-s(t)\,\delta _{jk}$ and
  $S_{ijk}=\dpp-\tilde s(t)\,\epsilon_{ijk}$ and we assume
 two equations of state:
\bb s(t)=\dpp w_s\,\rho (t),\qq \tilde s(t)=\dpp w_{\tilde s}\,\rho (t). \ee
Then the generalised Friedmann equations, i.e. the $tt$ and the $xx$ components of Einstein's equations, and Cartan's equations read:
\bb 3\,\frac{b^2-f^2}{a^2}\, &=&\Lambda +8\pi G\rho ,\label{tt}\\[2mm]
2\,\frac{b'}{a}\, +\,\frac{b^2-f^2}{a^2}\, &= &\Lambda - 8\pi  G p ,\label{xx}\\[2mm]
2\,\frac{a'-b}{a}\, &=&8\pi Gw_s\rho \label{cartan1}\\[2mm]
2\,\frac{f}{a}\, &=&8\pi Gw_{\tilde s}\rho \label{cartan2}.\ee

These equations agree with results in references \cite{symt,gmh} and in reference \cite{bloo} except for a missing 1/3 in front of the last term on the right-hand side 
 of equation (23) there. Note that this factor re-appears correctly in the subsequent equation (24).
 
 However we disagree with a result in reference \cite{mink} stating that the field equations imply that the function $f(t)$ (in our notations) must be constant. This result is reproduced in reference \cite{torLam}, probably because of a missing  factor 3 in its equation (20) (in the arXiv version) and it is only with this factor missing that the presumably constant $f$ can be interpreted as a cosmological constant.
 
Putting the pressure  to zero, $p=0$,
we have four equations for four unknown functions: $a,\,b,\,f$ and $\rho $. Equations (\ref{tt}) and (\ref{cartan2}) are algebraic, the other two equations, (\ref{xx}) and (\ref{cartan1}), are first order differential equations for $a$ and $b$. We use the algebraic ones to eliminate $\rho $ and $f $. Then we have a unique solution with two inital conditions $a(0)=a_0$ and $b(0)=b_0$. We therefore have five parameters, $a_0,\, b_0,\,\Lambda ,\, w_s$ and $w_{\tilde s}$. (We assume Newton's constant known.) These five parameters then fix $\rho (0)$ by use of equation (\ref{tt}):
\bb
1=\Omega _{m0}+\Omega _{\Lambda0} + 2\Omega _{s0} - \Omega _{s0}^2 +\,\frac{9}{4}\, \Omega_{\tilde s0}^2, \label{friedman} \ee
 with familiar dimensionless quantities:
\bb
\Omega _m\dpp=\,\frac{8\pi G\rho }{3H^2}\, ,\qq
\Omega _\Lambda \dpp=\,\frac{\Lambda  }{3H^2}\,, \qq
\Omega _s\dpp= w_sH\,\frac{8\pi G\rho }{2H^2}\,,
\qq
\Omega _{\tilde s}\dpp= w_{\tilde s}H\,\frac{8\pi G\rho }{3H^2}\,.
\ee
In particular, we see that the scale factor today $a_0$ has dropped out. This is well-known for  cosmology with vanishing spatial curvature  and remains true in presence of non-vanishing torsion. Note also that the sign of $f$ does not matter because only its square appears in the two Einstein equations. Therefore we may assume the state parameter $w_{\tilde s}$ to be non-negative.
 
 \section{Hubble diagram}
 
 To compute the Hubble diagram, we must solve the geodesic equations for co-moving galaxies and for photons \cite{berry}. For both, torsion decouples and they reduce to the geodesic equations with the Christoffel connection of the metric. Consequently the redshift is still given by $z=a_0/a(t)\,-1$ and the apparent luminosity $\ell$ is still related to the absolute luminosity of the standard candle $L$ by
 \bb \ell(t)= \,\frac{L}{4 \pi a^2_0\,x(t)^2}\, \,\frac{a(t)^2}{a^2_0}\, .\label{luminosity} \ee
We have put the earth at the origin of the Cartesian coordinates and the supernova on the $x$-axis:
\bb x(t):=\int_t^{t_0}\,\frac{\de\tilde t}{a(\tilde t)}\, .\ee

Note that the Einstein equations in presence of half-integer spin do feel torsion. However the link between the Hubble constant $H_0$ and $\de (z^2\ell )/\de z (0)$ is purely kinematical and therefore does not depend on torsion. This fact will be crucial to identify consistently the initial conditions of Friedmann's equations.

\section{Data analysis}

The data analysis used in this paper has been fully described in \cite{ts}. Only a brief reminder is given here. The type 1a supernovae Hubble diagram is constructed using the Union 2 sample \cite{union2} with 557 supernovae and the full systematic error matrix. The magnitude of supernovae is written as $M(z) = m_s + 2.5 \log \ell(z)$ where $m_s$ is a normalization parameter fitted to the data and $\ell(z)$ the apparent luminosity (\ref{luminosity}).

The apparent luminosity is computed using the generalized Friedmann equations (\ref{tt}), (\ref{xx}), (\ref{cartan1}) and (\ref{cartan2}). These equations are solved numerically by using the Runge-Kutta algorithm \cite{runge}.

The MINUIT \cite{minuit} package is used to fit the best cosmology by minimizing the $\chi^2$ defined as:
\bb \chi^2 = \Delta M^T V^{-1} \Delta M, \ee
where $\Delta M$ is the vector of differences between measured and expected magnitudes and $V$ the full covariance matrix including systematic errors.

Marginalization over unwanted parameters as $m_s$ and error estimates or contour constructions are obtained using the frequentist prescription \cite{pdg}. The Einstein-Cartan cosmology fit is performed with 3 or 4 free parameters ($m_s,\, \Omega_m, \,  \Omega_s \, \Omega_{\tilde s}$) while $\Omega_\Lambda$ is derived from the Friedmann-like equation (\ref{friedman}).

Table 1 presents the results of the fit of Einstein-Cartan's theory (parity even
and/or odd) and, for comparison, the results of the fit of the pure Einstein theory in the first line.
Because of very high non-Gaussianity, errors are given at 1 and 2 sigma level.
Minimum $\chi^2$ for all theories are statistically equivalent. If in the
parity even case the preferred value for $\Omega_m$ is compatible at a level of one
sigma with baryon matter density, in the odd-parity case, the preferred value of
$\Omega_m$  is in agreement with the total matter density of 0.27 published
by the WMAP collaboration  \cite{wmap}. This is not surprising since the
preferred value of $\Omega_{\tilde s}$ is exactly zero implying a flat Universe
in the pure Einstein theory. In all cases, the cosmological constant energy density is only slightly changed.

\begin{table}[htbp]
\begin{center}
\label{results}
\begin{tabular}{|c|c|c|c|c|c|} \hline
                 & $\Omega_m$              & $\Omega_\Lambda$         &       $\Omega_s$      &       $\Omega_{\tilde s}$   &$\chi^2_{min}$ \\ \hline
Einstein          & $0.35^{+0.10+0.15}_{-0.11-0.17}$  & $0.88^{+0.19+0.28}_{-0.11-0.32}$  &    $0.$    & $0.$    &$530.0$  \\ \hline
even-parity torsion   & $0.09^{+0.30+0.47}_{-0.07-0.08}$  & $0.83^{+0.10+0.12}_{-0.16-0.23}$  &  $0.04^{+0.01+0.02}_{-0.07-0.12}$ &$0.$ &$530.4$  \\ \hline
odd-parity torsion   & $0.27^{+0.03+0.06}_{-0.02-0.27}$  & $0.73^{+0.04+0.06}_{-0.11-0.32}$  & $0.$ &  $0.^{+0.22+0.55}_{-0.22-0.55}$ &$530.4$ \\ \hline
odd-even parity   & $0.08^{+0.27+0.9}_{-0.07-0.08}$  &$0.85^{+0.10+0.15}_{-0.15-0.25}$  & $0.04^{+0.02+0.06}_{-0.06-0.34}$ &$0.^{+0.01+0.6}_{-0.01-0.6}$ &$530.0$ \\ \hline
odd parity no $\Lambda$ & $0.01^{+0.02+0.03}_{-0.02-0.03}$  & $0.$  &$0.$ &  $0.66^{+0.01+0.02}_{-0.01-0.02}$ &$560.7$ \\ \hline
\end{tabular}
\caption[]{Fit results (1 and 2$\sigma$ errors) for Einstein and Einstein-Cartan theories with even and odd parity. No flatness constraint is imposed in the pure Einstein's theory.}
\end{center}
\end{table}

\begin{figure}[h]
\hspace{0.2cm}
\includegraphics[width=16cm, height=16cm]{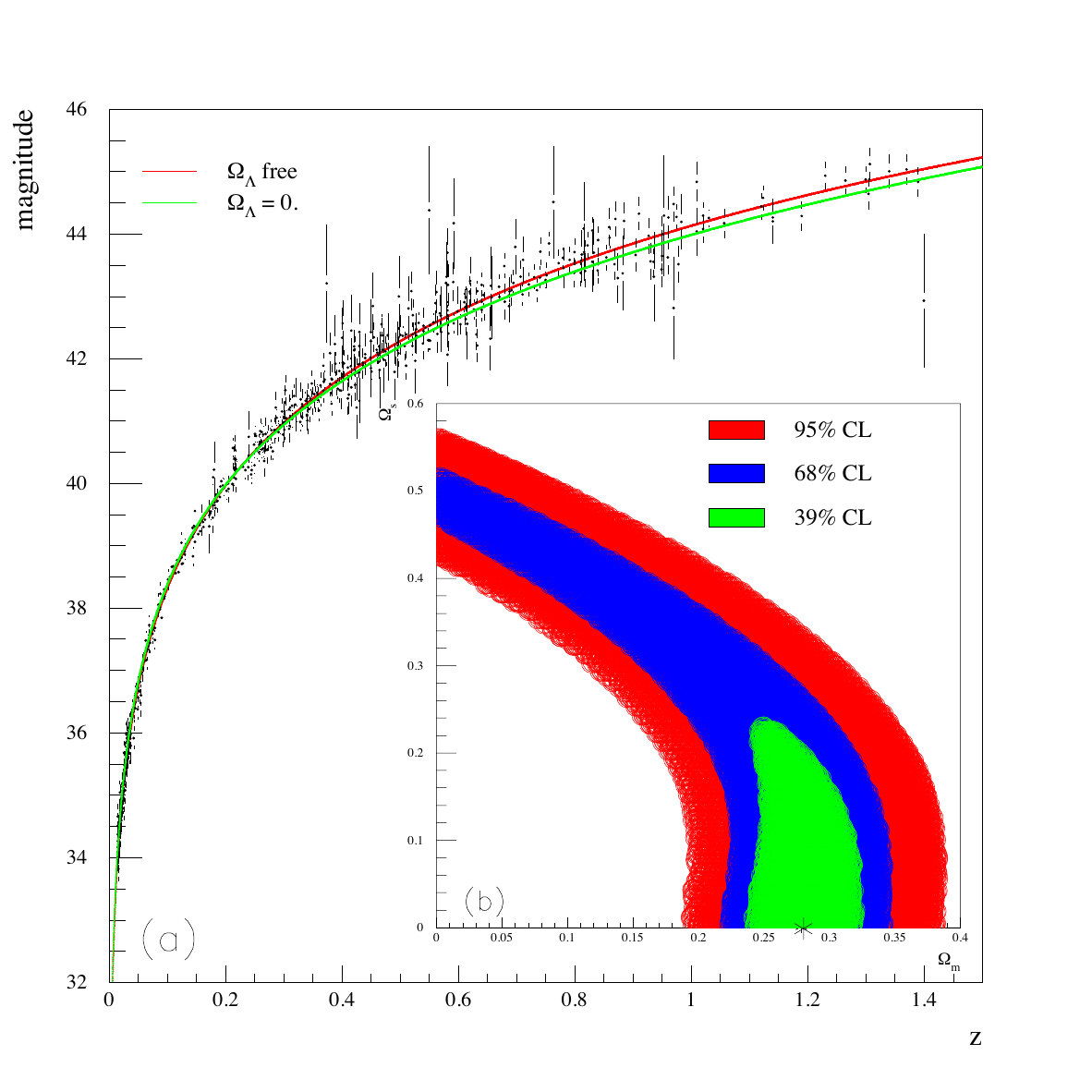}
\caption[]{\small (a) Fit results using the Union 2 Supernovae sample. The
  red (upper) curve corresponds to the Einstein-Cartan 3-fit ($m_s,\Omega_m,\Omega_{\tilde s}$) 
  while the green (lower) curve represents the 2-fit assuming a vanishing
  cosmological constant. (b) $39\%,68\%$ and $95\%$ confidence level contour
  in the ($\Omega_m,\Omega_{\tilde s}$) plane for the Einstein-Cartan 3-fit.}
\label{fig1}
\end{figure}

In figure \ref{fig1}(a) the result of the Hubble diagram fit with odd
Einstein-Cartan theory is shown (upper curve) with  data points and error
bars. As in the case of parity even Einstein-Cartan theory, the agreement
between fitted curve and data points is excellent. The lower curve shows the
fit resulting from putting the cosmological constant to zero for parity odd torsion. The agreement with
data points seems good and suggests that the cosmological constant can be replaced by parity odd torsion.

To test this hypothesis quantitatively, we use the log likelihood ratio technique. The log likelihood ratio is defined as:
\bb R = -2 Ln \frac{Sup(\llll (m_s,\Omega_{\tilde s},\Omega_\Lambda=0))}{ {Sup(\llll (m_s,\Omega_{\tilde s},\Omega_\Lambda))}}. \ee
Here  $Sup$ denotes the supremum of the likelihood function defined in term of the $\chi^2$ :
\bb \llll = \frac{1}{(2 \pi)^{(n/2)}|V|^{1/2}} e^{(-\frac{\chi^2}{2})}, \ee
$n$ is the number of data points and $V$ the full error matrix.
Thus the log likelihood ratio reads simply:
\bb R = \chi^2_{min,1}-\chi^2_{min,2}\,. \ee
The probability distribution of this test variable is approximately a $\chi^2$ distribution with degree of freedom equal to the difference between the degrees of freedom of both models, one in this case.

The minimum $\chi^2$ for the null cosmological constant hypothesis is equal to 560.7 while the minimum $\chi^2$ for odd parity torsion with cosmological constant is 530.4 (Table 1). The $p$-value is found to be equal to $6 10^{-8}$  corresponding to 5.4 $\sigma$ significance. 

Because systematic errors of supernovae intrinsic magnitude variations at high redshift (above 1) can be important, we check the null cosmological constant hypothesis using only supernovae at redshift below 1. The log likelihood ratio is found  to be 25 leading to a $p$-value of $5.3 10^{-7}$ or a significance of 5.01 sigma. Thus, the null cosmological constant hypothesis within the parity odd Einstein-Cartan theory is ruled out at more than 5 $\sigma$. 

For completeness, we perform the same analysis using simultaneously even and
odd parity torsion. The resulting fit is slightly better because of one
more fitted parameter. The $\chi^2$ of the fit is equal to 560.1 leading to
a $p$-value of $5.96 10^{-7}$ corresponding to a $5.4 \sigma$ significance.

\section{Conclusion}

We find that a fit of Einstein-Cartan's theory to the Hubble diagram is incompatible at 5 $\sigma $ level with the replacement of the cosmological constant by torsion, parity preserving or not. We think that the contrary claim in reference \cite{torLam} relies on a wrong coefficient in the field equations \cite{mink,torLam}.

\end{document}